\documentclass[preprint2]{aastex6}

\usepackage{lineno}
\usepackage{graphicx}

\begin{document}


\title{Modeling the Optical to Ultraviolet Polarimetric Variability from Thomson Scattering
in Colliding Wind Binaries}



\author{Richard Ignace}
\affil{Department of Physics \& Astronomy,
East Tennessee State University,
Johnson City, TN 37614, USA}

\author{Andrew Fullard}
\affil{Department of Physics and Astronomy,
Michigan State University,
East Lansing, MI 48824, USA}

\author{Manisha Shrestha}
\affil{Astrophysics Research Institute, Liverpool John Moores University, Liverpool Science Park, 146 Brownlow Hill, \\
Liverpool L3 5RF, UK}

\author{Ya\"el Naz\'e\altaffilmark{1}}
\altaffiltext{1}{FNRS Senior Research associate}
\affil{GAPHE, UR STAR, Universit\'e de Li\`ege, All\'ee du 6 Ao\^ut 19c (B5C), B-4000 Sart Tilman, Li\`ege, Belgium}

\author{Kenneth Gayley}
\affil{Department of Physics \& Astronomy, University of Iowa, Iowa City, IA 52242, USA}

\author{Jennifer L. Hoffman}
\affil{Department of Physics \& Astronomy,
University of Denver, 2112 E. Wesley Ave., Denver, CO 80208, USA}

\author{Jamie R. Lomax}
\affil{Department of Physics,
United States Naval Academy,
572C Holloway Rd.,
Annapolis, MD 21402, USA}

\author{Nicole St-Louis}
\affil{D\'epartement de physique, Universit\'e de Montr\'eal, Complexe des Sciences, 1375 Avenue Th\'er\`ese-Lavoie-Roux, Montr\'eal (Qc), H2V 0B3, Canada}

\begin{abstract}

Massive star binaries are critical laboratories for measuring masses
and stellar wind mass-loss rates.  A major challenge is inferring
viewing inclination and extracting information about the colliding
wind interaction (CWI) region.  Polarimetric variability from
electron scattering in the highly ionized winds provides important
diagnostic information about system geometry.  We combine for the
first time the well-known generalized treatment of
\citet{brown_polarisation_1978} for variable polarization from
binaries with the semi-analytic solution for the geometry and surface
density CWI shock interface between the winds based on
\citet{canto_exact_1996}.  Our calculations include some simplifications
in the form of inverse square-law wind densities and the assumption
of axisymmetry, but in so doing arrive at several robust conclusions.
One is that when the winds are nearly equal (e.g., O\,+\,O binaries),
the polarization has a relatively mild decline with binary separation.
Another is that despite Thomson scattering being a gray opacity,
the continuum polarization can show chromatic effects at ultraviolet
wavelengths but will be mostly constant at longer wavelengths.
Finally, when one wind dominates the other, as for example in
WR\,+\,OB binaries, the polarization is expected to be larger at
wavelengths where the OB component is more luminous, and generally
smaller at wavelengths where the WR component is more luminous.
This behavior arises because from the perspective of the WR star,
the distortion of the scattering envelope from spherical is a minor
perturbation situated far from the WR star.  By contrast, the
polarization contribution from the OB star is dominated by the
geometry of the CWI shock.

\end{abstract}



\section{Introduction} \label{sec:intro}
Despite comprising the rarest stellar mass group, massive stars ($>$ 8 M$_\odot$) are the most important originators of elements in the Universe because they synthesize and distribute heavy elements when they explode as supernovae \citep{2013ARA&A..51..457N}. Massive stars also enrich the interstellar medium during their pre-supernova lifetime via their strong stellar winds.  High levels of mass loss also affects the evolution of massive stars, in particular the nature of their remnants
\citep{2008A&ARv..16..209P,2012ARA&A..50..107L,2014ARA&A..52..487S}.

Most massive stars spend a large fraction of their lives in binary systems with other massive stars; approximately 50\% are thought to engage in mass exchange with a close companion \citep{sana_binary_2012}. Interactions between companions drive the evolutionary paths that can shape both stars' subsequent fates \citep[e.g.,][]{2012ARA&A..50..107L,2016A&A...585A.120S}.

Colliding-wind binaries can teach us a great deal about the individual stars and their winds because the geometry of the interaction region is  dependent on the relative mass-loss rates and velocities of the binary components. 
Early theories describing these binaries used momentum flux \citep[e.g.][]{girard_winds_1987} and ram-pressure balance \citep[e.g.][]{kallrath_dynamics_1991}, or hydrodynamic models \citep[e.g.][]{stevens_colliding_1992}, and even some semi-analytic work \citep[e.g.][]{2007ApJ...655.1002P,2009ApJ...703...89G,1992ApJ...389..635U,canto_exact_1996}. Further work in the area has focused on hydrodynamic simulations \citep[][]{parkin_3d_2008,lamberts_high-resolution_2011,macleod_hydrodynamic_2020} and predictions of line profiles in both optical \citep{luhrs,georgiev_line_2004,ignace_modelling_2009} and X-rays \citep{2003MNRAS.346..773H,mossoux16,mossoux21}. The result of these models, combined with a number of phase-resolved observations, is that we have generally a good understanding of the expected geometry of colliding winds \citep[e.g.][]{rau4220,gosset30a,sana152248b,gosset22,wil2009,ken2010,fah2011,naze09,cazo2014,rau228766,2015A&A...573A..43L,gosset21a,naze5980,callingham_two_2020, rodriguez2020}.

Given that massive star winds are strongly ionized, it is natural to consider Thomson scattering as the dominant scattering opacity in the winds, which in turn can polarize the observed light. The resulting polarization is sensitive to the geometry of the scattering regions. The classic \citet[][hereafter BME]{brown_polarisation_1978} model approximates the time-varying continuum polarization caused by the illumination of circumstellar material in a binary system viewed at an arbitrary inclination angle. Those authors assume the electron scattering region is optically thin. Their approach allows for a general geometry, but for binary stars they consider two point sources of illumination and a co-rotating scattering envelope.

\citet{brown_effect_1982} extended the BME model to consider elliptical orbits, and \citet{fox_stellar_1994} further extended the formalism to consider finite-size illuminators. \citet{fox_stellar_1994} showed that occultation is only important in very close binary systems, where the separation of stars is less than 10 times the radius of the primary. However, none of these enhancements to the theory specifically addressed the effects of colliding winds in the time-dependent polarization results. Furthermore, the effects of the wind collision regions on the wavelength dependence of polarization have not been considered as part of this theoretical framework. However, such polarized signals associated with colliding winds have been observed in several systems \citep{st._-louis_polarization_1993, lomax_v444_2015}. A modelling effort in this domain is therefore critically important.

Polarization models of stellar wind bow shock structures produced by the interaction of stellar winds with a local ambient medium shows that significant polarization can arise from scattering of light in these structures \citep{Shrestha_2018, Shrestha_2021}. Modeling the polarization signal caused by colliding wind geometries has been done specifically for the system V444 Cyg \citep{st._-louis_polarization_1993,kurosawa_mass-loss_2002}, but a general formalism has not yet been produced. 
In this paper we derive a consistent model for the polarization signal produced by wind collision regions in massive binary systems. In Section~\ref{sec:model} we describe our model of polarimetric variability from optically thin electron scattering in a shock illuminated by two stars. In Section~\ref{sec:results} we derive expressions for the polarization signal based on the system parameters, and show how these expressions lead to chromatic and orbital effects in polarization. We summarize our results in Section~\ref{sec:summary}.

\section{Polarimetric Variability from Thin Electron Scattering}\label{sec:model}

Our treatment is based on that of BME, who
presented a thorough theoretical construction for thin electron scattering in
a generalized envelope with an arbitrary
number of illuminating point sources.  In regards to our application for a binary system, we assume the colliding wind interaction (CWI) is axisymmetric about the line of centers (LOC) joining the two stars (i.e., we ignore the Coriolis effect in our example cases, though this has been detected in at least one colliding-wind binary; \citealt{lomax_v444_2015}).  We further assume that the separate winds of the two stars are each spherically symmetric up to the CWI.  As a result, in the notation of BME, we have $\gamma_1=\gamma_2=\gamma_3=\gamma_4=0$,
and the only factors that are nonzero are $\gamma_0$ and $\tau_0$.  In our development we will modify this notation slightly.

BME then considered the more
limited scenario of a binary system with a circular orbit and corotating
envelope.  Our approach allows for elliptical orbits (which \citealt{brown_effect_1982} later
considered), and for the shape and density of the bowshock, we employ
the analytical solution of \citet{canto_exact_1996}.  

The \cite{canto_exact_1996} solution is predicated on strong radiative cooling.
There are two initial concerns about adopting this model.  The first is that
radiative cooling leads to thin shell instabilities \citep[e.g.,][]{2011MNRAS.418.2618L}.
The second is that radiative cooling is limited to relatively close binaries,
orbital periods of order a week \citep[e.g.,][]{2004ApJ...611..434A}.
These issues have significant relevance for predicting X-ray spectra, where
the temperature distirbution along the shock is important; or when simulating
emission line profile shapes, where the detailed vector velocity field is crucial.
However, our case deals with electron scattering and continuum polarization.  Unlike the case of X-ray diagnostics where the distribution of hot gas is important, we can safely assume the gas 
is everywhere highly ionized for computing scattering polarization.  Of chief importance to our case is that the
\cite{canto_exact_1996} derivation is conveniently analytic and driven by considerations of ram pressure balance which captures much of the key physics.
Most, but not all, of our examples involve either equal winds or binaries with one dominant wind, and our general conclusions based on the \cite{canto_exact_1996}
model are fairly robust.

As in the BME formalism, our approach assumes point source illumination.  This is reasonable
when the binary separation is not too small (of order the stellar radii).
One distinction, however, is that we do account for finite stellar size
when evaluating volume integrals.  This is not incompatible with BME; we are merely explicit about its inclusion. 

A final point about our use of the BME treatment is that due to the axisymmetry, we employ
the notation and approach of BME for a single star and use superposition
in our application to a binary system.  In doing so, our notation departs
from BME, although we still employ similar variables.  The following
sections introduce our geometry for the systems, our application
of BME to axisymmetric binaries, and semi-analytic solutions for
the variable polarization based on the \citet{canto_exact_1996} solution for the CWI.

\subsection{Geometry and Stellar Properties}

In our model, the winds of the two stars and the intervening colliding
wind shock are prescribed using primarily polar coordinates for
each star.  We define the primary star to be the one with the stronger wind
in terms of momentum flux, $\dot{M}v_\infty$, where $\dot{M}$
is the mass-loss rate and $v_\infty$ is the terminal wind speed.
The secondary is then the weaker wind case in terms of this
product.  We typically use subscripts ``1'' and ``2'' to identify
 primary and secondary.

We introduce spherical coordinates with respect to the axis that is the line
of centers between the two stars.  Coordinates centered on the primary star 
$(r_1,\theta_1,\phi_1)$ are such that $\theta_1 = 0$ in the direction
of the secondary.  Likewise, the coordinates linked to the secondary star $(r_2,\theta_2,\phi_2)$ also have $\theta_2 = 0$ in the direction
of the primary.  Frequently, our approach employs the standard cosine
notation, $\mu_1 = \cos \theta_1$ and $\mu_2 = \cos \theta_2$.

The individual winds are taken
to be spherical with densities varying with the inverse square of the distance from the star.  Consequently, we are ignoring the wind acceleration zone that is relevant whenever the bow shock forms close to either or both stars.  We do not include the potential for radiative
braking of the stronger wind if it enters the region of dominance of 
the companion stellar flux \citep{1997ApJ...475..786G,lomax_v444_2015}, nor the possibility that the
stronger wind might in some cases ram directly into the photosphere of the secondary star.

With spherical winds and constant speed radial outflow, the density relations for the primary and secondary winds are:

\begin{eqnarray}
\rho_1 & = & \frac{\dot{M}_1}{4\pi\,r_1^2\,v_1} = \rho_{1,0}\,\left(\frac{R_{1,\ast}}{r_1}\right)^2, \label{eq:rho1} \\
\rho_2 & = & \frac{\dot{M}_2}{4\pi\,r_2^2\,v_2} = \rho_{2,0}\,\left(\frac{R_{2,\ast}}{r_2}\right)^2,~{\rm and} \label{eq:rho2} \\
n_1 & = &  n_{1,0}\,\left(\frac{R_{1,\ast}}{r_1}\right)^2, \label{eq:n1} \\
n_2 & = &  n_{2,0}\,\left(\frac{R_{2,\ast}}{r_2}\right)^2,
\end{eqnarray}

\noindent where $n_{\rm j}$ for $j=1,2$ are electron number densities
with corresponding scale parameter $n_{\rm j,0} = \rho_0/\mu_{\rm
e}m_H$, for $m_H$ the mass of hydrogen and $\mu_{\rm e}$ the mean
molecular weight per free electron.  Since we consider terminal
speed flow, we have omitted the subscript $\infty$ for the primary
and secondary wind speeds in equations~(\ref{eq:rho1}) and
(\ref{eq:rho2}).

\subsection{Bow Shock Model}

\begin{figure}
\plotone{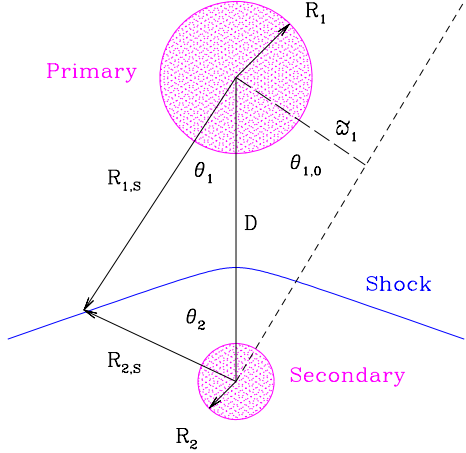}
\caption{Top-down illustration of the two stars, primary and secondary in magenta and
the bowshock, in blue, formed by the colliding winds.  Labeled variables are defined
in text.
\label{fig1}}
\end{figure}

To explore polarimetric variability from colliding wind shocks, we
choose the formulation of \citet{canto_exact_1996} for their semi-analytic bow shock
solution involving radiative cooling and
two stellar spherical winds at terminal speed.  This solution
specifically assumes radiative cooling, and the bow shock takes the
form of an interface of infinitesimal thickness between the two otherwise spherical
winds.  This interface for the CWI is axisymmetric and characterized with a surface density.

Here we reproduce in brief the key expressions for the solution of
\citet{canto_exact_1996}, but with a few changes of notation.  \citet{canto_exact_1996} use
a ``1'' subscript for the primary, but no subscript for the secondary;
we use a ``2'' subscript for the secondary.  \citet{canto_exact_1996} use $\sigma$
for the surface mass density, while we use $\Sigma = \sigma / \mu_{\rm
e}m_H$ for the surface number density of electrons.  In principle
$\mu_{\rm e}$ is a function of coordinate in either wind and in the
bow shock itself; however, for simplicity, we assume $\mu_{\rm e}$
is constant throughout the shock.

The bow shock geometry and its surface density are related to
two fundamental ratios:

\begin{eqnarray}
\beta & = & \frac{\dot{M}_2v_2}{\dot{M}_1v_1},~{\rm and} \\
\alpha & = & v_2/v_1.
\end{eqnarray}

\noindent The first of these, $\beta\le 1$, is the ratio of wind momentum of the secondary compared to the primary; the second , $\alpha$, is  the ratio of wind terminal speeds.
Figure~\ref{fig1} provides a schematic of the binary
system with intervening CWI region between the two stars.
The radial distances of the bow shock from the stars
are denoted as $R_{1,S}$ and $R_{2,S}$, with

\begin{eqnarray}
R_{2,S} = D\,\frac{\sin \theta_1}{\sin(\theta_1+\theta_2)},~{\rm and} \\
R_{1,S} = \sqrt{D^2+R^2_{2,S}-2\,D\,R_{2,S}\,\cos\theta_1},
\end{eqnarray}

\noindent where $D$ is the separation between the two stars at any moment.
We denote the standoff radii of the bowshock from each star along the line of centers as $R_{1,0}$ and $R_{2,0}$, with

\begin{eqnarray}
R_{2,0} = \frac{\beta^{1/2}}{1-\beta^{1/2}}\,D,~{\rm and} \\
R_{1,0} = \frac{1}{1-\beta^{1/2}}\,D.
\end{eqnarray}

While the above relations are analytic formulations in
$\theta_1$ and $\theta_2$, the relation between the
two coordinate angles is implicit, making the solution overall
semi-analytic.  The relationship of the angles is given by,

\begin{equation}
\frac{\theta_1}{\tan \theta_1} = 1 + \beta\left(\frac{\theta_2}{\tan\theta_2}
	-1\right).
\end{equation}

\noindent Note that the asymptotic angles (``opening angles'') for the bow shock
are given by

\begin{eqnarray}
\theta_{2,\infty}-\tan\theta_{2,\infty} = \frac{\pi}{1-\beta},~{\rm and}\\
\theta_{1,\infty}=\pi - \theta_{2,\infty}.
\end{eqnarray}

\noindent The case $\beta=1$ corresponds to a planar shock between
identical stars and winds, with $\theta_{1,\infty}=\theta_{2,\infty}=\pi/2$.

The final key ingredient for modeling the polarimetric variability is
the surface number density distribution.  Again from \citet{canto_exact_1996}, this is
given by

\begin{eqnarray}
\frac{\Sigma}{\Sigma_0} & = & \frac{\sin(\theta_1+\theta_2)}{\sin \theta_1
	\sin \theta_2} \times \\ \nonumber
	& & \left[\beta(1-\cos\theta_2)+\alpha(1-\cos\theta_1)\right]^2 \times \\ \nonumber
	& & \left\{ \left[ \beta(\theta_2-\sin\theta_2\cos\theta_2 +
	(\theta_1-\sin\theta_1\sin\theta_2)\right]^2  \right. \\ \nonumber
	& & \left. + \left( \beta\sin^2\theta_2-\sin^2\theta_1\right)^2 \right\}^{-1/2}
\end{eqnarray}

\noindent where the scaling constant is

\begin{equation}
\Sigma_0 = \frac{\dot{M}_2/\mu_{\rm e}m_H}{2\pi\beta\,D\,v_2}
	= \frac{2R_{2,\ast}}{\beta\,D}\,R_{2,\ast}n_{2,0}= \alpha\,
	\frac{2R_{1,\ast}}{D}\,R_{1,\ast}n_{1,0}.
	\label{eq:Sigma0}
\end{equation}

\noindent The last step in the above expression represents the surface
number density as being twice the column depth of the wind of
the secondary from the bow shock to infinity.

In the special case of $\beta=1$ for a planar shock forming from
two identical stars, $R_{1,S} = R_{2,S} = D/2\mu$, where $\mu=\cos
\theta$ with $\theta=\theta_1=\theta_2$.  Combined with $\alpha=1$,
the surface density then simplifies to

\begin{equation}
\frac{\Sigma}{\Sigma_0} = 4\,\frac{\cos\theta\,(1-\cos\theta)^2}
	{\sin\theta(\theta-\sin\theta\cos\theta)}.
	\label{eq:planarSigma}
\end{equation}

Having defined the geometry of the CWI interface and its properties,
we turn next to characterizing the electron scattering polarization.

\subsection{Thin Scattering}

As previously noted, we employ the approach of BME in application to 
the results of \citet{canto_exact_1996}, which is explicitly
axisymmetric.  In the treatment
of BME, ignoring absorption and assuming that the total amount of
scattered light is small compared to the specific luminosities of
either star, the polarization is given by

\begin{equation}
p_{\rm tot} = \tau\,(1-3\gamma)\,\sin^2 i=p_0\,\sin^2 i,
    \label{eq:B&M}
\end{equation}

\noindent where $p_{\rm tot}$ is the polarization, $\tau$ is an angle-averaged
optical depth of the envelope, $\gamma$ is called the shape factor,
and $i$ is the viewing inclination relative to the symmetry axis
of the envelope.  We introduce $p_0$ for conveniently representing the
product of optical depth and envelope shape.
The definitions of $\tau$ and $\gamma$ are

\begin{equation}
\tau = \frac{3}{16}\sigma_T\,\int\int n(r,\mu)\,dr\,d\mu,
\end{equation}

\noindent and

\begin{equation}
\gamma = \frac{\int\int n(r,\mu)\,dr\,\mu^2\,d\mu}{\int\int n(r,\mu)
	\,dr\,d\mu},
\end{equation}

\noindent where $n(r,\mu)$ is the axisymmetric distribution of electrons
throughout the scattering volume.  If the scattering region is spherically symmetric, $\gamma=1/3$, and the polarization
is zero.  For a wind whose density varies as the inverse of the squared distance, $\tau = 3\tau_\ast/8$,
with $\tau_\ast = n_0\,\sigma_T\,R_\ast$ being the radial optical depth of the wind in electron scattering.

The integrals for $\tau$ and $\gamma$ are defined with respect to the star
center, and the treatment for axisymmetry does not require top-down
symmetry.  Thus one can introduce $\tau_1$ and $\gamma_1$ associated
with polarization from scattering of starlight from the primary,
and then $\tau_2$ and $\gamma_2$ for scattering of starlight by
the secondary.  Being optically thin, the results add linearly
as weighted by the wavelength-dependent luminosities of the two stars:

\begin{equation}
p_0 = \frac{L_1(\lambda)\,p_1+L_2(\lambda)\,p_2}{L_1(\lambda)+L_2(\lambda)}.
    \label{eq:ptot}
\end{equation}

\noindent This result, in our notation, is equivalent to equations~(6a)
and (7) from BME for axisymmetry.  The dependence on viewing
inclination is implicit in equation~(\ref{eq:ptot}) via 
equation~(\ref{eq:B&M}).  As pointed out by BME, we have that $i_1=i_2=i$ so
that $p_{\rm tot} = p_0 \sin^2 i$ as in equation~(\ref{eq:B&M}) earlier.

Note that for a particular geometry as expressed by the
wind and orbital properties, $p_1$ and $p_2$ have generally different values
but are defined with respect to the same axis, the LOC between the stars.  
While these values are not wavelength-dependent (i.e., chromatic), $p_0$ can
be chromatic because the two illuminating sources will generally have
different spectral energy distributions (SEDs).  

This last point deserves additional comment.  When one star dominates the brightness of the system in a given wavelength range, the polarization
will be flat and take the polarization value of the dominant star. At wavelengths for which the emission of both stars follows a Rayleigh-Jeans law, the relative contribution to the total luminosity of the stars will always be the same, hence $p_{\rm tot}$ will also be flat. However, for stars of unequal temperatures, at wavelengths around the Wien peak, the ratio of specific luminosities will vary, and either $p_1$ or $p_2$ may dominate, or the dominant terms may switch.  Thus for hot stars, $p_0$ will be chromatic at short wavelengths, despite the fact that electron scattering is gray.

\section{Model Results}\label{sec:results}

\subsection{Expressions for the Polarization}

Recall that for BME, the polarization depends on source parameters
$\tau$ and $\gamma$, but these in turn depend on angle-averaged
column densities of free electrons.  It is convenient to
introduce two varieties of angle-averaged column densities, 
one that is a zeroth-order moment and one that is a second-order moment:

\begin{equation}
\langle N \rangle = \frac{1}{2}\int\int n(r,\mu)\,dr\,d\mu,
\end{equation}

\noindent and

\begin{equation}
\langle \tilde{N} \rangle = \frac{1}{2}\int\int n(r,\mu)\,dr\,\mu^2\,d\mu.
\end{equation}

\noindent Then 

\begin{equation}
\tau = \frac{3}{8}\sigma_T\,\langle N \rangle,
\end{equation}

\noindent and

\begin{equation}
\gamma = \langle \tilde{N} \rangle/\langle N \rangle,
\end{equation}

\noindent where these parameters would have subscripts 1 or 2 for the primary or secondary stars, respectively.  For example, the primary wind with density given by equation~(\ref{eq:n1}) would have a first-moment column density of $\langle N_1 \rangle = n_{1,0}R_{1,\ast}$.

\begin{figure}[t]
\plotone{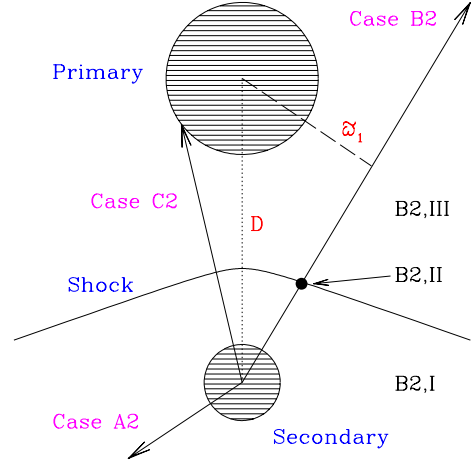}
\caption{Similar to Fig.~\ref{fig1}, the two stars and colliding
wind shock are illustrated.  For calculation of the polarization
properties, column density moments involve three different ray
categories, here identified as A2, B2, and C2 in relation to
the secondary star.  The dotted line is the line of centers (LOC)
for separation $D$ between the stars.  The impact parameter
$\varpi_1$ is the same as for Fig~\ref{fig1}.  For Case B2, the contributions
associated with I (path segment), II (point), and III (semi-bound
line segment) are identified.
\label{figrays}}
\end{figure}

In application to the colliding wind binary, there are 3 key angular
regimes to consider in relation to each of the two stars.  These 3 solid angle
sectors contribute to the
parameters $\gamma_{1,2}$ and $\tau_{1,2}$.  These will be detailed
next in terms of Cases A, B, and C, with reference to Figure~\ref{figrays}.

Figure~\ref{figrays} is a copy of Figure~\ref{fig1} in terms of the
stars and the CWI (labeled as ``Shock''), now used to emphasize the 3 angular regimes.  There is still the LOC with separation $D$.  Using the secondary as example, three rays are labeled with A2, B2, and C2.  Similar rays could be drawn as originating from the primary, which would then be labeled A1, B1, and C1.
Consequently,

\begin{equation}
    \langle N \rangle = \langle N_A \rangle + \langle N_B \rangle + \langle N_C \rangle,
\end{equation}

\noindent and

\begin{equation}
    \langle \tilde{N} \rangle = \langle \tilde{N}_A \rangle + \langle \tilde{N}_B \rangle + \langle \tilde{N}_C\rangle.
\end{equation}

\subsubsection{Case A:  Rays not intersecting the colliding wind shock}
\label{sub:caseA}

For $\theta_1>\theta_{1,\infty}$ or $\theta_2>\theta_{2,\infty}$,
the ray does not intersect the shock nor the opposite star, but travels
strictly through its own wind.  In this case the angle-averaged column
densities are trivial for an inverse square law, with

\begin{eqnarray}
\langle N_{A1} \rangle & = & \frac{1}{2}n_{1,0}R_{1,\ast}\,(1+\mu_{1,\infty}), \\
\langle \tilde{N}_{A1} \rangle & = & \frac{1}{6}n_{1,0}R_{1,\ast}\,(1+\mu^3_{1,\infty}), \\
\langle N_{A2} \rangle & = & \frac{1}{2}n_{2,0}R_{2,\ast}\,(1+\mu_{2,\infty}), \\
\langle \tilde{N}_{A2} \rangle & = & \frac{1}{6}n_{2,0}R_{2,\ast}\,(1+\mu^3_{2,\infty}).
\end{eqnarray}

\subsubsection{Case B:  Rays intersecting the shock but not the opposite star}
\label{sub:caseB}

When a ray intersects the shock, there are three distinct contributions
to the relevant angle-averaged column densities:  a segment of
the star's own wind, the surface density at the shock itself, and
finally a segment through the wind of the opposite star.
The ray does not intercept the surface of the opposite
star, a case treated in the next section.

We denote these three contributions as I for own wind, II
for the shock, and III for the opposite wind, and use these
as subscripts to accompany 1 and 2 to signify whether the rays originate
from the primary or the secondary star.  Note that the angular
integrations 
are affected by the ray missing or intercepting the opposite star.

The first segment is the column within the star's own wind from
its surface to the shock.  The columns are:

\begin{eqnarray}
\langle N_{B1,I} \rangle & = & \frac{1}{2}n_{1,0}R_{1,\ast}\,\int_{\mu_{1,\infty}}^{\mu_{2,\ast}}\,\left[1-\frac{R_{1,\ast}}{R_{1,S}(\mu)}\right]\,d\mu, \label{eq:B1I} \\
\langle N_{B2,I} \rangle & = & \frac{1}{2}n_{2,0}R_{2,\ast}\,\int_{\mu_{2,\infty}}^{\mu_{1,\ast}}\,\left[1-\frac{R_{2,\ast}}{R_{2,S}(\mu)}\right]\,d\mu, \\
\langle \tilde{N}_{B1,I} \rangle & = & \frac{1}{2}n_{1,0}R_{1,\ast}\,\int_{\mu_{1,\infty}}^{\mu_{2,\ast}}\,\left[1-\frac{R_{1,\ast}}{R_{1,S}(\mu)}\right]\,\mu^2\,d\mu, \\
\langle \tilde{N}_{B2,I} \rangle & = & \frac{1}{2}n_{2,0}R_{2,\ast}\,\int_{\mu_{2,\infty}}^{\mu_{1,\ast}}\,\left[1-\frac{R_{2,\ast}}{R_{2,S}(\mu)}\right]\,\mu^2\,d\mu, \label{eq:B2I}
\end{eqnarray}

\noindent where $\mu_{1,\ast}=\cos\theta_{1,\ast}= \sqrt{1- (R_{1,\ast}/D)^2}$ and
$\mu_{2,\ast}=\cos\theta_{2,\ast} = \sqrt{1-(R_{2,\ast}/D)^2}$.
In the integrand, the
fraction being subtracted represents the missing column
in the star's own wind that would be present if not for the shock.

\begin{figure}[t]
\plotone{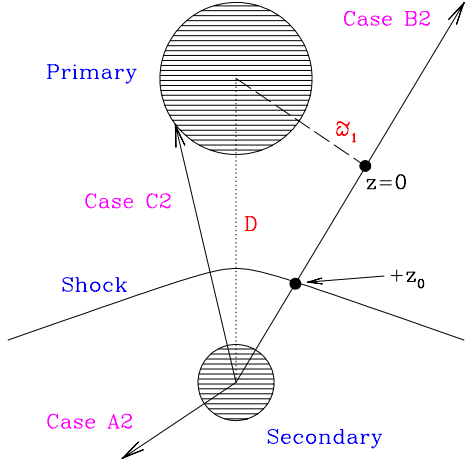}
\caption{Similar to Fig.~\ref{figrays}, the two stars and colliding
wind shock are illustrated.  The cases A2, B2, and C2 in relation to
the secondary star are indicated.  The $z$ coordinate is indicated in this version,
with $z_0$ labeled.  For columns evaluated along radials from the secondary, $z$ is defined in the wind of the primary; and vice versa.
\label{figz}}
\end{figure}

At the shock itself, the surface density contributes to the column
along the ray, owing to the accumulation of material entering the
shock and outflowing along its surface.  However, the kinematics are
not of relevance here; only the surface density itself matters.  What is relevant
is the projection of the surface with respect to the ray.  The
contribution to the column depends on a direction cosine between
the radial unit vector and the local normal to the surface.  We
represent these as $K_1(\mu)$ and $K_2(\mu)$.  Omitting the subscripts,
$K$ for each star is given by

\begin{equation}
K(\mu) = \frac{1}{\sqrt{1 + \left(\frac{1}{R_S}\frac{dR_S}{d\theta}\right)^2}}.
\end{equation}

\noindent Then the contribution by the shock to the columns becomes

\begin{eqnarray}
\langle N_{B1,II} \rangle & = & \frac{1}{2}n_{1,0}R_{1,\ast}\,\int_{\mu_{1,\infty}}^{\mu_{2,\ast}}\,K_1(\mu)\,\Sigma(\mu)\,d\mu, \label{eq:A} \\
\langle N_{B2,II} \rangle & = & \frac{1}{2}\,\int_{\mu_{2,\infty}}^{\mu_{1,\ast}}\,K_2(\mu)\,\Sigma(\mu)\,d\mu, \label{eq:B} \\
\langle \tilde{N}_{B1,II} \rangle & = & \frac{1}{2}\,\int_{\mu_{1,\infty}}^{\mu_{2,\ast}}\,K_1(\mu)\,\Sigma(\mu)\,\mu^2\,d\mu, \label{eq:C} \\
\langle \tilde{N}_{B2,II} \rangle & = & \frac{1}{2}\,\int_{\mu_{2,\infty}}^{\mu_{1,\ast}}\,K_2(\mu)\,\Sigma(\mu)\,\mu^2\,d\mu. \label{eq:D}
\end{eqnarray}

The third region is in the wind of the opposite star, beyond the shock.  The integration in
radius depends on whether the ray strikes the star or not.  The former case
is treated in the next section.  Here we express the columns along the
ray from the shock that stretches to infinity with impact parameter
$\varpi$.  Consider, for example, a ray originating from the secondary
at orientation $\theta_2$.  The ray intercepts the shock at $R_{2,S}$.
Further, the ray is in the wind of the primary.  The impact parameter
for that ray is $\varpi_1 = D\sin\theta_2$.

For a spherical wind with inverse square density, the integration
for the column along such a chord is analytic:

\begin{equation}
\int_{-\infty}^{z_0} n_0\,\frac{R_\ast^2}{r^2}\,dz = n_0\,R_\ast\,\frac{R_\ast}
	{\varpi(\mu)}\,\int_0^{\theta_0}\,d\theta = 
	n_0\,R_\ast\,\frac{R_\ast}{\varpi(\mu)}\,\theta_0(\mu),
	\label{eq:theta0}
\end{equation}

\noindent where the subscripts 1 and 2 have been suppressed for
this general result.  Take again the example of a ray from the
secondary.  Then $z_0$ refers to the $z$-coordinate in the wind of
the primary corresponding to $R_{1,S}$ and $\varpi_1$. Figure~\ref{figz} shows the location of $z_0$ in relation to the system components. With
$\theta_0$ defined with respect to the stellar axis, one can show that $\theta_0
+\theta_1 + \theta_2 = \pi$. 

The next step is to integrate in $\mu$.  Again using
the secondary as an example, this integration will be of
the form $\theta_0(\mu_2)d\mu_2/\varpi_1 \sim \theta_0(\theta_2)d\theta_2$.
The end result from the columns for section III is:

\begin{eqnarray}
\langle N_{B1,III} \rangle & = & \frac{1}{2}n_{2,0}R_{2,\ast}\,\frac{R_{2,\ast}}{D}\\ \nonumber
& & \times\int_{\theta_{1,\infty}}^{\theta_{2,\ast}}\,(\pi-\theta_1-\theta_2)\,d\theta_1, \\
\langle N_{B2,III} \rangle & = & \frac{1}{2}n_{1,0}R_{1,\ast}\,\frac{R_{1,\ast}}{D}\\ \nonumber
& & \times\int_{\theta_{2,\infty}}^{\theta_{1,\ast}}\,(\pi-\theta_1-\theta_2)\,d\theta_2, \\
\langle \tilde{N}_{B1,III} \rangle & = & \frac{1}{2}n_{2,0}R_{2,\ast}\,\frac{R_{2,\ast}}{D}\\ \nonumber
& & \times\int_{\theta_{1,\infty}}^{\theta_{2,\ast}}\,(\pi-\theta_1-\theta_2)\,\cos^2\theta_1\,d\theta_1, \\
\langle \tilde{N}_{B2,III} \rangle & = & \frac{1}{2}n_{1,0}R_{1,\ast}\,\frac{R_{1,\ast}}{D}\\ \nonumber
& & \times\int_{\theta_{2,\infty}}^{\theta_{1,\ast}}\,(\pi-\theta_1-\theta_2)\,\cos^2\theta_2\,d\theta_2.
\end{eqnarray}

\noindent where $\sin\theta_{1,\ast} = R_{1,\ast}/D$ and
$\sin\theta_{2,\ast} = R_{2,\ast}/D$.  When the ray intercepts the
opposite star, the lower limit to the integral in
equation~(\ref{eq:theta0}) is no longer 0 in $d\theta$.  Thus
the integrands for the angular integrations above are different over
the solid angle extent of the opposite star.

For Case B, one adds the contributions from the different segments to
obtain

\begin{eqnarray}
\langle N_{B1} \rangle & = & \langle N_{B1,I}\rangle + \langle N_{B1,II}\rangle + \langle N_{B1,III}\rangle , \\
\langle N_{B2} \rangle & = & \langle N_{B2,I}\rangle + \langle N_{B2,II}\rangle + \langle N_{B2,III}\rangle , \\
\langle \tilde{N}_{B1} \rangle & = & \langle \tilde{N}_{B1,I}\rangle + \langle \tilde{N}_{B1,II}\rangle + \langle \tilde{N}_{B1,III}\rangle , \\
\langle \tilde{N}_{B2} \rangle & = & \langle \tilde{N}_{B2,I}\rangle + \langle \tilde{N}_{B2,II}\rangle + \langle \tilde{N}_{B2,III}\rangle .
\end{eqnarray}

\vspace{11pt}

\begin{figure}
\plotone{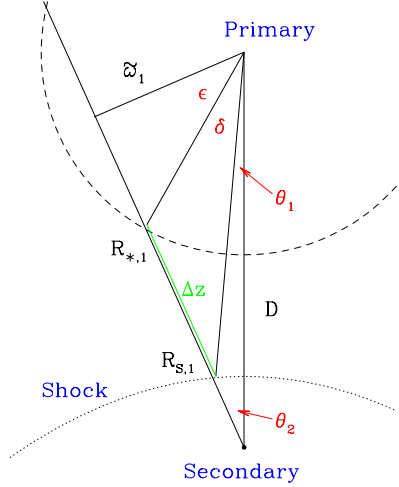}
\caption{Similar to Fig.~\ref{fig1}, but zoomed to emphasize angular quantities
used for case C calculations.  The primary star is shown at top with its extent
a dashed curve.  The shock is a dotted curve.  For the secondary star at bottom, only its center
is indicated.
\label{figangles}}
\end{figure}

\begin{figure}[t]
\plotone{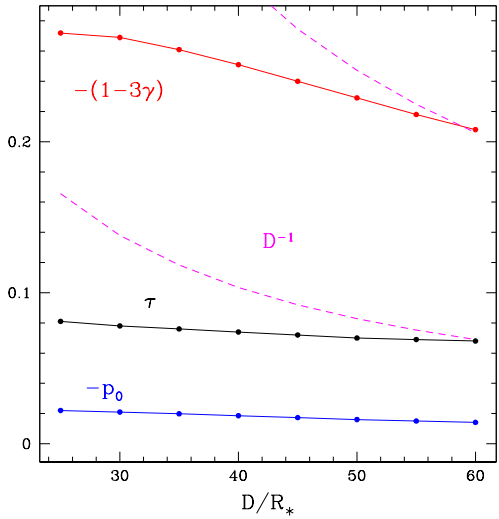}
\caption{For the case of two equal stars and a planar shock,
$\tau=\tau_1=\tau_2$ and $\gamma=\gamma_1=\gamma_2$.  The figure
shows the variations of these parameters along with $p_{\rm tot}$
as a function of binary separation, $D/R_\ast$, where $R_\ast = R_1
=R_2$.  Example lines of $D^{-1}$ variation are shown for reference;
the parameters determining the colliding wind polarization decline
much more slowly than $D^{-1}$.  Note that since $p_{\rm tot}<0$, its
negative is plotted for convenience.}
\label{fig3}
\end{figure}

\begin{table*}[t]
\caption{Stellar and Wind Properties for Parameter Study Displayed in 
Figure~\ref{fig4}}
\label{tab1}
    \centering
\begin{tabular}{cccccccc}
\hline \hline Panel & $R_{2,\ast}$ & $R_{1,\ast}$ & $\dot{M}_2$ &
$\dot{M}_1$ & $v_2$ & $v_1$ & $D$ \\
& $(R_\odot)$ & $(R_\odot)$ & $(10^{-7} M_\odot~{\rm yr}^{-1})$ &
$(10^{-7} M_\odot~{\rm yr}^{-1})$ & (10$^3$ km s$^{-1}$) & (10$^3$ km s$^{-1}$)
& $(R_\odot)$ \\ \hline
 Top &  10 &  10 &  65 & 100 & 2 & 2 & 100  \\     
 Top &  10 &  10 &  70 & 100 & 2 & 2 & 100 \\    
 Top &  10 &  10 &  75 & 100 & 2 & 2 & 100 \\    
 Top &  10 &  10 &  80 & 100 & 2 & 2 & 100 \\    
 Top &  10 &  10 &  85 & 100 & 2 & 2 & 100 \\    
 Top &  10 &  10 &  90 & 100 & 2 & 2 & 100 \\    
 Top &  10 &  10 &  95 & 100 & 2 & 2 & 100 \\
 Top &  10 &  10 & 100 & 100 & 2 & 2 & 100 \\
     &     &     &     &     &   &   & \\
 Mid &  10 &  10 & 100 & 100 & 1.3 & 2 & 100\\
 Mid &  10 &  10 & 100 & 100 & 1.4 & 2 & 100 \\    
 Mid &  10 &  10 & 100 & 100 & 1.5 & 2 & 100 \\    
 Mid &  10 &  10 & 100 & 100 & 1.6 & 2 & 100 \\    
 Mid &  10 &  10 & 100 & 100 & 1.7 & 2 & 100 \\    
 Mid &  10 &  10 & 100 & 100 & 1.8 & 2 & 100 \\    
 Mid &  10 &  10 & 100 & 100 & 1.9 & 2 & 100 \\    
 Mid &  10 &  10 & 100 & 100 & 2.0 & 2 & 100 \\    
     &     &     &     &     &   &   & \\
 Bot &   3 &  10 & 100 & 100 & 2 & 2 & 100   \\    
 Bot &   4 &  10 & 100 & 100 & 2 & 2 & 100  \\   
 Bot &   5 &  10 & 100 & 100 & 2 & 2 & 100  \\   
 Bot &   6 &  10 & 100 & 100 & 2 & 2 & 100  \\   
 Bot &   7 &  10 & 100 & 100 & 2 & 2 & 100  \\   
 Bot &   8 &  10 & 100 & 100 & 2 & 2 & 100  \\   
 Bot &   9 &  10 & 100 & 100 & 2 & 2 & 100  \\   
 Bot &  10 &  10 & 100 & 100 & 2 & 2 & 100  \\ \hline
\end{tabular}
\end{table*}

\subsubsection{Case C:  Rays that intercept the opposite star}
\label{sub:caseC}

In the final scenario,  case C, the ray intersects the opposite
star.  This modifies the upper limit for the integral associated with segment
III.  Segments I and II from case B are the same for case C; only segment III differs.
To be explicit, for segment C,I for a star's own wind up to the shock, one still
uses equations~(\ref{eq:B1I})--(\ref{eq:B2I}) for B,I but with different limits to the integrals.  The lower limit is $\mu_\ast$ for star 1 or 2 as appropriate, and the upper limit is $+1$ for all the integrals.  Similarly, the contribution C,II at the shock uses equations~(\ref{eq:A})--(\ref{eq:D}) for B,II also with $\mu_\ast$ (again, as appropriate for 1 or 2) for the lower limit and $+1$ for the upper limit.  It is only C,III that requires reconsideration, as follows.

One way of expressing case C,III is that when $\varpi<R_\ast$, or
alternatively when $\theta<\theta_\ast$, for primary or
secondary as the case may be, the
radial integral along the chord is still given by an angle,
but this angle is not $\theta_0$.  Instead we introduce new angles
$\epsilon$ and $\delta$, with

\begin{equation}
\theta_1+\theta_2+\epsilon+\delta = \pi/2.
\end{equation}

\noindent  Figure~\ref{figangles} shows the locations of $\epsilon$ and $\delta$ with respect to the system. We have that $\varpi_1 = D \sin\theta_2 = R_{1,\ast}
\cos \epsilon_1$, so that $\epsilon_1$ is defined in terms of
$\theta_2$ for a ray originating from the secondary and intercepting
the primary.  When the case is reversed, all the subscripts
are reversed.  This allows us to find $\delta$ from the definition above,
and the radial integration along the ray segment is $\delta$. 

The columns now become

\begin{eqnarray}
\langle N_{C1,III} \rangle & = & \frac{1}{2}n_{2,0}R_{2,\ast}\,\frac{R_{2,\ast}}{D}\int_0^{\theta_{2,\ast}}\,\delta(\theta_1)\,d\theta_1, \\
\langle N_{C2,III} \rangle & = & \frac{1}{2}n_{1,0}R_{1,\ast}\,\frac{R_{1,\ast}}{D}\int_0^{\theta_{1,\ast}}\,\delta(\theta_2)\,d\theta_2, \\
\langle \tilde{N}_{C1,III} \rangle & = & \frac{1}{2}n_{2,0}R_{2,\ast}\,\frac{R_{2,\ast}}{D}\int_0^{\theta_{2,\ast}}\,\delta(\theta_1)\,\cos^2\theta_1\,d\theta_1, \\
\langle \tilde{N}_{C2,III} \rangle & = & \frac{1}{2}n_{1,0}R_{1,\ast}\,\frac{R_{1,\ast}}{D}\int_0^{\theta_{1,\ast}}\,\delta(\theta_2)\,\cos^2\theta_2\,d\theta_2.
\end{eqnarray}

\vspace{11pt}

\subsection{Special Case of a Planar Shock}
\label{sub:planar}

For a binary consisting of two identical stars, $\alpha=\beta=1$, and the
resulting CWI is a planar shock located midway between the stars at
$D/2$ from either one.  Appendix~\ref{AppA} details the simplifications
that result for this scenario, in particular an analytic expression
for $R_{S,1}=R_{S,2}$ in terms of $\mu$, $K_1=K_2$, and
the surface density distribution.

We introduce the simplifying notation $\tau=\tau_1=\tau_2$ and $\gamma=\gamma_1=\gamma_2$,
and display in Figure~\ref{fig3} how these properties vary with
separation between the stars.  Note also that $p_1=p_2\equiv p_0$, and
because the two stars have $L_1=L_2$, $p_0$ is a constant at all wavelengths.  For this figure we assume a stellar radius of $10R_\ast$, wind speed $2000$ km s$^{-1}$,
and mass loss rates $10^{-5} M_\odot$ yr$^{-1}$.  The polarization amplitude scales with $\dot{M}/v$
for the case of equal stars.

By our convention for Figure~\ref{fig1},
both $(1-3\gamma)$ and $p_0$ are negative, so are shown in Figure~\ref{fig3}
as multiplied by $-1$ for convenience.  Since many, but not all, terms associated with the
calculation scale as $D^{-1}$, dotted magenta curves are shown with that scaling for comparison.  The outcome is that
the polarization amplitude $p_0$ does indeed decline with binary separation, but much
less steeply than $D^{-1}$; its behavior is closer to linear for the chosen parameters and distances shown.

\subsection{Parameter Study}

\begin{figure}
\plotone{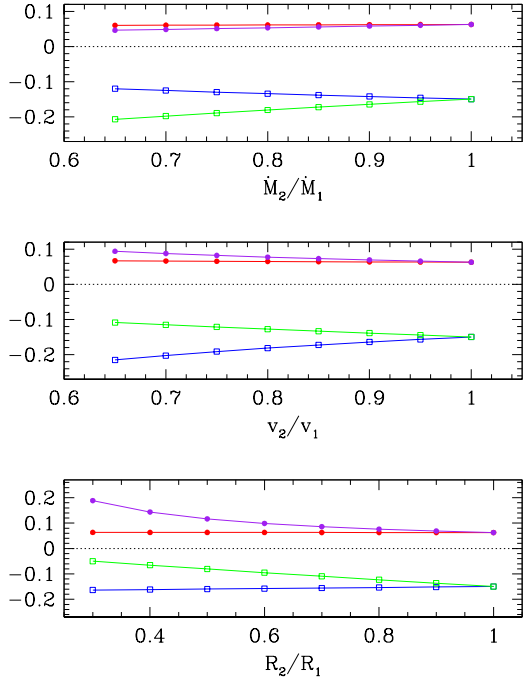}
\caption{Variation of optical depth $\tau_1$ (red) and $\tau_2$
(purple), shape factors $(1-3\gamma_1)$ (blue), and $(1-3\gamma_2)$
(green) for the parameter study with model properties detailed in
Table~\ref{tab1}; solid points are optical depths and open squares
are for the shape factors involving $\gamma$.  The top panel
shows how the polarization varies with the ratio of mass-loss rates
for the stars; middle is for the ratio of wind speeds, $v_2/v_1 =
\alpha$; bottom is for the ratio of stellar radii.  For each panel
(``top'', ``mid'', ``bot''), the 8 points on the curves correspond
to model calculations for the 8 parameter combinations shown in
Table~\ref{tab1}.
\label{fig4}}
\end{figure}

We conducted a parameter study of relatively similar winds for primary
and secondary components.  Results are displayed in Figure~\ref{fig4}, with
model parameters identified in Table~\ref{tab1}.  The figure has 3 panels
-- top, middle (``mid''), and bottom (``bot'') -- with model parameters similarly
grouped in Table~\ref{tab1}.  
The top is for variation of the ratio of
mass-loss rates (which turns out to be $\beta$ because $\alpha$ is fixed); the middle is 
for variation of the ratio of the wind terminal speeds (which is $\alpha = v_2/v_1$, but also
$\beta=\alpha$ for fixed mass-loss rates); and bottom is for 
variation of radii.  Note that for the top
and middle panels, the far right side corresponds to equal winds and a planar shock.
For the bottom panel, $\beta=1$ and the shock is always planar.

In the top panel, the ratio of mass-loss rates varies from
0.65 up to 1.0.  The four curves are for $\tau_1$ in red, $\tau_2$ in
purple, $1-3\gamma_1$ in blue, and $1-3\gamma_2$ in green.
While the primary wind has the higher optical depth, the deviation of the envelope
from spherical increases faster for the secondary (green) than the primary (blue)
as $\beta$ declines.  This is an important feature of the discussion in
Section~\ref{sec:wr_ob_binaries}, where small $\beta$ values are emphasized 
as being typical of WR+OB binary systems.

For the middle panel, $\beta=\alpha$, yet the behavior is reversed.  Lowering
the wind speed of the secondary actually elevates the density scale for
its wind.

Finally for the bottom panel, $\beta=\alpha=1$ is fixed, and the wind shock is
planar.  Despite the geometry being invariant, the polarization depends on the radii of
the two stars.  This arises because
 the polarization properties scale with column density,
which are generally inverse to radius.  For example, with a secondary smaller than the
primary yet having the same mass-loss rate and wind speed, the column density is higher in the secondary wind, so $\tau_2$ (purple) increases with decreasing $R_2$.

\subsection{Chromatic Effects}
\label{sub:chromatic}

To illustrate chromatic effects, we fixed a particular set of binary parameters
while allowing the temperatures of the two stars to vary.  The fixed properties lead to $p_1=0.009$ and $p_2 = 0.192$.
In this example, the binary components are separated by $40~R_\odot$.  
The secondary
is twice as large ($10R_\odot$) as the primary ($5R_\odot$); the two winds have equal speeds (2000 km s$^{-1}$); and the mass-loss
rate for the primary is $3.3\times$ larger than the secondary ($\dot{M}_2 = 3\times 10^{-6}\,M_\odot$ yr$^{-1}$).  Additionally,
we took $\sin i = 1$; the effect of viewing inclination is to scale the curves by $\sin^2 i$ at
all wavelengths.

\begin{figure}
\plotone{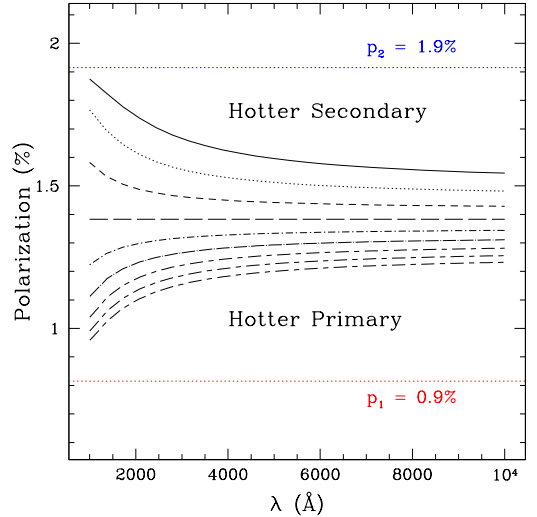}
\caption{Variation of the polarized continuum with wavelength, here shown from the FUV
to 1~micron.  The stars are treated as Planckian.  The temperature of the secondary is
fixed at 25,000~K, and the temperature of the primary varies from 16,000~K to 40,000~K
in 3,000~K increments.  The particulars of the stellar and wind parameters for this
illustrative case are described in Sect.~\ref{sub:chromatic}.  For the selected parameters, the limiting polarizations $p_1$ and $p_2$ are indicated with horizontal red and blue dotted lines,
respectively.
\label{fig5}}
\end{figure}

To illustrate changes in the spectropolarimetric continuum shape, we treated the two stars as simple Planckian sources with effective temperatures $T_1$ and $T_2$.
We fixed the temperature of the secondary at $T_2=25,000$~K.  We varied the temperature
of the primary from $16,000$~K to $40,000$~K in $3,000$~K
intervals, and display the results in Figure~\ref{fig5}, where polarization is
shown as positive.  These variations may not be consistent with actual combinations
of parameters for real stars.  The point of the exercise is to highlight
the fact that when the stars have different spectral energy distributions, the
continuum polarization is not generally constant with wavelength, even though electron scattering
is gray.  Only when the two stars have equal temperatures is the continuum polarization
truly flat at all wavelengths.

Note especially that the polarization signal changes strongly from the FUV through the optical to 1 micron.
For massive stars with typical temperatures well in excess of $10,000$~K,
both the primary and secondary spectra are
in the Rayleigh-Jeans
tail in the optical, and so the continuum will always be flat or nearly flat in that waveband.  It is only in the
UV that the polarization deviates significantly from constant.  For the selected parameters, the polarization actually drops toward the UV when the primary is hotter (i.e., more luminous), owing to the fact that $p_1 \ll p_2$.  By contrast, when the secondary is hotter (i.e., more luminous), the polarization increases significantly.  Ultimately, for any combination of binary parameters,
when the more luminous star in the UV also has the higher polarimetric component (i.e., $p_1$ or $p_2$), the polarization is enhanced in the UV relative to the optical; when the more luminous UV source has the lower polarimetric component, the polarization will drop toward the UV.

The behavior in Figure~\ref{fig5} is included in the formalism of
BME, but is specific to binaries with two hot stars.
Other categories of binaries can certainly show rather different behavior. For example, consider symbiotic stars, which involve a hot white dwarf and a cool giant
star \citep[e.g.,][]{1991A&A...248..458M}.  In such a case, the polarigenic opacity may be more complex than simple electron scattering (i.e., not simply gray opacity),
and may not involve a colliding wind but perhaps instead accretion onto a disk.  
Nonetheless, if the scattering opacity is dominated by Thomson scattering, the combination of an optically-bright component with a UV-bright component would
yield a wavelength-dependent polarization that would reveal a telltale gradual variation
from the FUV to the IR.

In any situation where intrinsic polarization from Thomson
scattering is observable over a wide spectral domain (separable from interstellar polarization
either by its binary variation in the time domain, or line effects in the stellar winds, or modeling
the wavelength dependence of the interstellar polarization), these results show that
the residual wavelength dependence of the intrinsic
polarization offers a unique and important diagnostic.
Via the following analysis, we obtain complementary leverage in our understanding of
both the different polarizations produced by the two stellar light sources, and also the spectral shape
of the continua of both stars.
This stems from the grayness of Thomson scattering, which implies that the sole source of wavelength dependence
in the intrinsic
polarization derives from the contrasting brightnesses of the two stars.
Hence if ${\cal L}_{21}(\lambda)$ is the wavelength-dependent ratio of the secondary brightness to the
primary, then the total polarization $q(\lambda)$ presents as a brightness-weighted
average of the wavelength independent
polarization induced by the
primary light source, $q_1$, and that induced by the secondary, $q_2$, according to
\begin{equation}
    q(\lambda)   =  \frac{q_1 \ + \ q_2 {\cal L}_{21}(\lambda) }{ 1 \ + \ {\cal L}_{21}(\lambda) } \ .
\end{equation}
If the spectral shape contrast ${\cal L}_{21}(\lambda)$ is regarded as known by the stellar spectral
types, then observing $q$ at two different wavelengths
that sample suitably different values of ${\cal L}_{21}(\lambda)$ allows the above equation to separate the
$q_1$ and $q_2$ contributions.
This separation of the polarizations caused by the two different light sources allows a unique probe of
the geometry of the wind collision zone.

Furthermore, to the extent that $q_1$ and $q_2$ are expected to be wavelength independent, a self-consistency
check on the assumed ${\cal L}_{21}(\lambda)$ becomes possible by inverting the above equation into
\begin{equation}
    {\cal L}_{21}(\lambda) = \frac{ q_1 \ - \ q(\lambda) }{q(\lambda) \ - \ q_2} \ .
\end{equation}
To whatever extent this inferred brightness contrast deviates from its assumed value, we have the 
opportunity to update it to recover consistency with the polarized spectrum $q(\lambda)$.
For example, a $(q_1 , q_2)$ pair can be inferred from wavelength pairs generated by fixing a wavelength
at the UV end of the observed $q(\lambda)$ and sweeping the second wavelength over the full observed
range.
If the assumed ${\cal L}_{21}(\lambda)$ contains errors, that would generate a curve in $(q_1 , q_2)$ space rather
than a single consistent point.  
Then by fixing the second wavelength at its longest value and sweeping the first wavelength back toward
the UV, the curve is closed back to its starting $(q_1 , q_2)$ point.
The resulting closed curve then gives an estimate of the preferred $(q_1 , q_2)$ value near the center
of this curve, and that preferred $(q_1 , q_2)$ pair then allows ${\cal L}_{21}(\lambda)$ to
be self-consistently updated via the above equation.
The wavelength independence of the intrinsic polarization contributions
$q_1$ and $q_2$, assuming they are dominated by Thomson scattering in 
the colliding winds, then provides an improved estimate of ${\cal L}_{21}(\lambda)$ and an independent
check of our understanding of binary spectral types.
Also, when both $q_1$ and $q_2$ are appreciable, independent knowledge of both allows an important
probe of the colliding wind geometry, since the two stars illuminate that geometry differently.

On the other hand, in situations where one contribution dominates, say $q_1$, as may be the case in WR/O 
binaries discussed below, the wavelength dependence of the intrinsic $q(\lambda)$ 
directly inherits the wavelength dependence of ${\cal L}_{21}(\lambda)$
via $q(\lambda) = q_1 / \left ( 1 + {\cal L}_{21}(\lambda) \right ) $.
Hence in this case we have an even more direct handle on the correct brightness contrast between
the two stars over the full wavelength regime of the observed polarization.
The wider that wavelength regime accessed by our technology, the more powerful is this constraint,
underscoring the value of extending our polarization capabilities into the FUV range for understanding
binaries containing hot stars.

\subsection{Orbital Effects}
\label{sub:orbital}

\begin{figure}
\centering
\includegraphics[width=0.9\columnwidth]{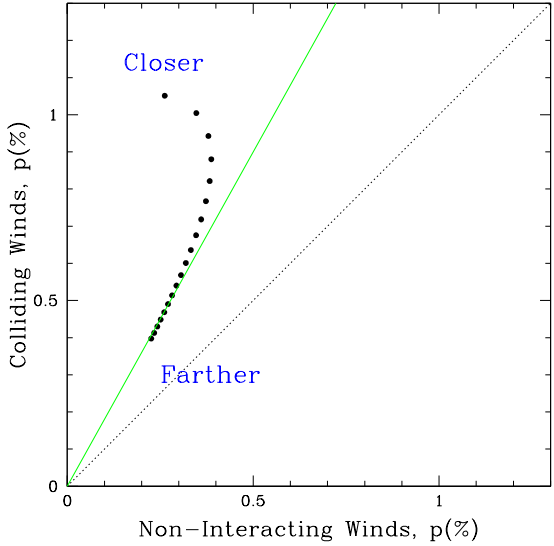}
\caption{Comparison of the percent polarization from a face-on binary system with colliding winds (CWs) against that from 
a binary with non-interacting winds (NIWs; \S~\ref{sub:orbital}).  The points represent separations of $25R_\odot$ to $115R_\odot$
in $5R\odot$ increments, with monotonic sequence as indicated by ``Closer'' and ``Farther''.  Wind
collision significantly increases the level of polarization of CWs over NIWs, but less so as the binary
separation increases to large values.  The green line represents the analytic derivation
from App~\ref{AppA} for wide-binary separations.}
\label{figorb}
\end{figure}

The polarimetric properties of the colliding wind system depend on the binary separation, $D$.  When the orbit is circular (i.e., $D$ is constant), the values $p_1$ and $p_2$ are constant as well.  For an eccentric orbit with eccentricity $e$ and semi-major axis $a$, the binary separation varies as

\begin{equation}
 D(\varphi) = a\,\frac{1-e^2}{1+e\cos\varphi},
\end{equation}

\noindent where $\varphi$ is the orbital azimuth, defined so that
$\varphi=0$ corresponds to periastron.  Thus $p_1$ and $p_2$ are
functions of orbital phase through the variation of $D(\varphi)$.

The polarization also varies throughout the orbital motion because the inclination, $i$, of the LOC between the stars changes relative to the observer's line of sight. Let $i_{\rm orb}$ be the viewing inclination of the orbital plane, so that $i_{\rm orb}=0^\circ$ is a top-down view of the orbit and $i_{\rm orb}=90^\circ$ is an edge-on view.  Despite a fixed orientation of the orbital plane,
$i_{\rm orb}$, our construction for calculating polarization depends on the system axis defined by the LOC between the stars, and this rotates in the fixed orbital plane to produce variability.

The time-variable polarization is given by

\begin{eqnarray}
q & = & p_0(D)\,\sin^2 i(t)\,\cos 2\psi(t),~{\rm and} \\
u & = & p_0(D)\,\sin^2 i(t)\,\sin 2\psi(t),
\end{eqnarray}

\noindent where $D=D(t)$ for eccentric orbits, $t$ depends on orbital phase through the azimuth ($\varphi$ of the LOC), and the polarization position angle $\psi$ relates to the orbital azimuth and the fixed viewing inclination of
the orbital plane  via

\begin{equation}
\tan \psi = -\cos i_{\rm orb}/\tan \varphi.
\end{equation}

\noindent The inclination, $i$, for the LOC to the viewer's line of sight is given by

\begin{equation}
\cos i = \sin i_{\rm orb}\,\cos \varphi.
\end{equation}

In order to understand the relevance of CWIs for the polarization level, we introduce the
idea of ``non-interacting winds'' (NIWs).  The concept of a NIW provides a reference against which
to compare the physical case of CWI polarization arising from the shock.  What CWI ultimately represents is a redistribution of matter from the two stellar winds via the wind collision.  There is polarization without a CWI, because each of the two stars shines on the wind of the other, even if both winds remain spherical.  The CWI represents another contribution by
 breaking spherical symmetry.  For wide binary separations ($D\gg R_\ast$), we expect 
 the two cases to become proportional, since the column
depths of the various regions will scale as $D^{-1}$ (see App.~\ref{AppA}).

\begin{figure}
\centering
\includegraphics[width=0.49\columnwidth]{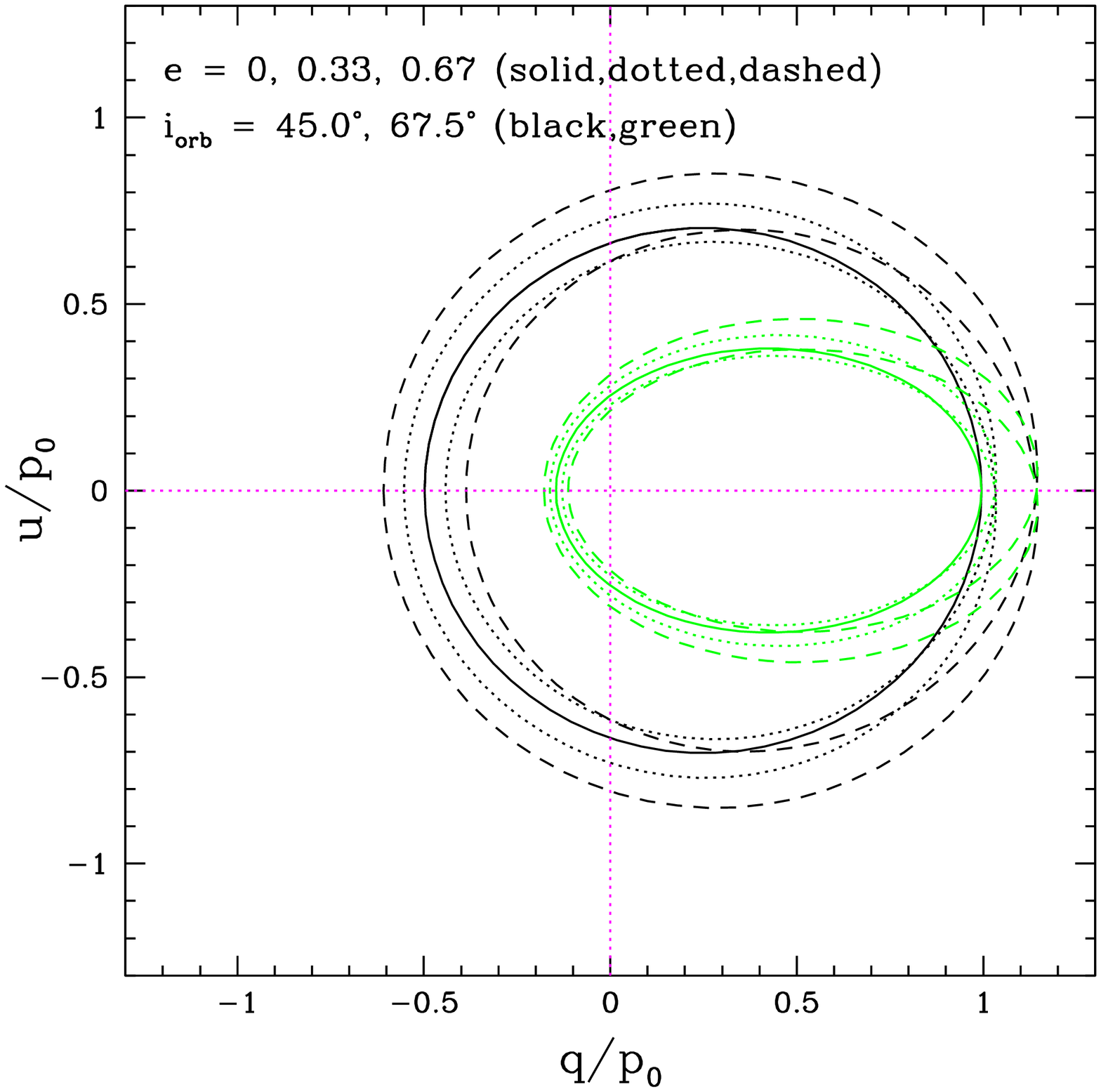}
\includegraphics[width=0.49\columnwidth]{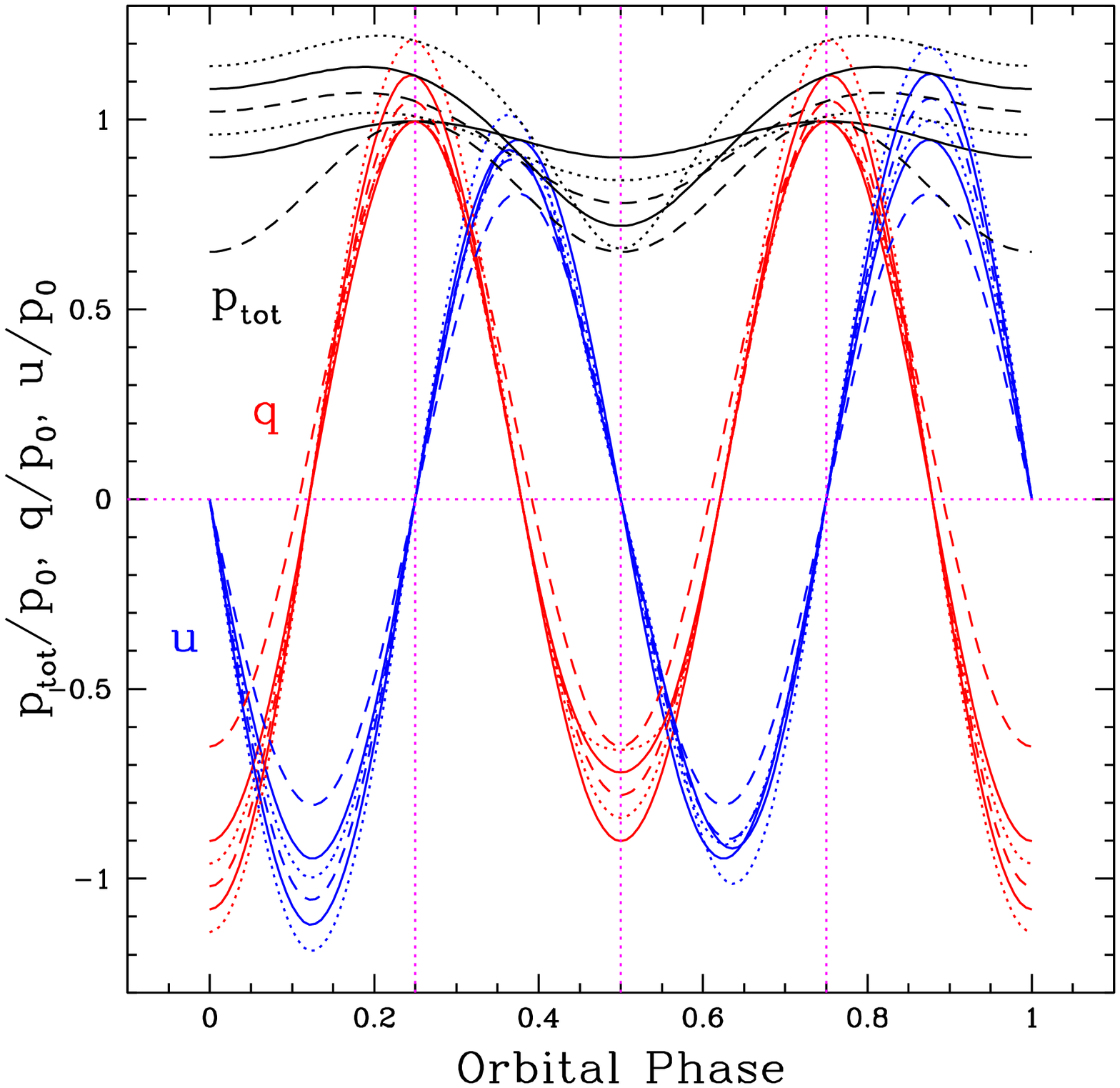}
\caption{(Top) Example $q-u$ loops for a binary consisting of identical stars, for which $p_1=p_2\equiv p_0$.  Eccentricities $e$ and inclinations $i_{\rm orb}$ are indicated for line type and color.  There are two loops per orbit, but with eccentric orbits, the pair separate into different sizes except for $e=0$.  (Bottom) Same 6 models as top, now displayed as lightcurves in $q$ (red), $u$ (blue), and total polarization $p_{\rm tot}$ (black).  Periastron passage is at phase 0.0, and apastron at phase 0.5.  The line types still relate to $e$ as in the top panel.  The case $i_{\rm orb} = 45^\circ$ has the higher polarization in $p_{\rm tot}$ and larger amplitude variation in $q$ and $u$.}
\label{fig6}
\end{figure}

We thus define a NIW simply as a superposition of the two separate binary winds as if no collision takes place and neither wind impacts the opposite star.  This means each wind is spherically 
symmetric about its own star, which contributes no polarization.  Instead, polarization arises
only from scattering of starlight from the secondary by the wind of the primary, and vice versa.
We do, however, account for occultation of the wind behind each respective star in calculating
the polarization.

Our results are shown in Figure~\ref{figorb}, with the polarization amplitude in percent for the NIW
along the horizontal, and for the CWI along the vertical.  For this example we assume two equal winds
and thus a planar shock for the CWI case.  The points represent stellar separations of $D=25\,R_\odot$ to
$115\,R_\odot$ in $5\,R_\odot$ increments, with closer and farther separations labeled for the sequence.
The values are for a face-on binary, oriented so that $u=0$ and $q<0$, and $p=| q|$ for this 
plot.  The dotted diagonal indicates where the polarizations would be equal; the solid green line is the asymptotic relation derived in Appendix~\ref{AppA}.

For the CWI case, we fitted a linear regression to $p_0(D)$ in the case of a binary with two identical stars (i.e., planar CWI shock) to obtain $p_0 = 1.328 - 0.0111D$, normalized so that $p_0 = 1\%$ at $D=30~R_\odot$.  For the NIWs, we used the same stellar wind and star properties as
for the wind collision case.  The polarization from a CWI is larger than for a NIW, often by a significant
factor.  At larger separations, the trend is for the points to approach the diagonal line,
signifying that the wind collision is becoming irrelevant.

We recognize that we are using a model for a bow shock with radiative cooling, and that
at large separations, the cooling will be adiabatic.  Even so, with identical stars and winds,
the shock will still be planar, and the qualitative conclusion remains valid even if the
quantitative values are inaccurate.  

Figure~\ref{fig6} displays a suite of polarimetric variations for CWIs with different values of $e$ and $i_{\rm orb}$.  We show model variable polarization curves for  inclinations 
and  eccentricities, as labeled.  The top panel displays the resulting $q-u$ loops; the bottom panel shows polarized light curves as a function of orbital phase.  
At $i_{\rm orb}=90^\circ$ (not shown), all curves in Figure~\ref{fig6} would become horizontal lines with only $q$ variation but no $u$ variation.  Note also that $p_{\rm tot}=\sqrt{q^2+u^2}$.

\subsection{Special Case of WR+OB Binaries}\label{sec:wr_ob_binaries}

Among the more extreme massive star colliding wind systems ($\beta \ll 1$) are the ones involving an evolved WR~star with an OB companion.  While the wind speeds of the two stars can be comparable in this case, the mass-loss rate of the WR~wind will be one to several orders of magnitude larger than for the OB component.  As a result, the CWI shock is significantly displaced from the WR star and considerably closer to the OB star; it also significantly confines the spatial scope of the OB wind.  On the other hand, the WR and OB components may or may not have comparable luminosities.  In terms of a UV study, the situation can be ideal for extracting information about the orbital parameters and properties of the CWI region from both temporal and chromatic effects.

The scenario of $\beta \ll 1$ offers some simplifications for the problem of the polarization.  Foremost is that the ``primary'' (defined as above as the WR~wind with higher mass loss, not necessarily the more luminous component) is relatively far from the CWI shock.  Consequently, one expects the angle-averaged column densities over the WR~wind component to approach
zero.  The associated column densities for the secondary wind evaluated at the 
primary will be small.  It may seem that $p_1 = p_{WR}$ would be dominated by
the CWI shock, but this may not mean that the WR~component dominates the polarization, since the CWI is relatively far removed, and thus only acts as a perturbation on the otherwise spherical wind of the WR~star.  The result for the O~star, $p_2 = p_O$, is less clear.  Its wind has lower column density than the WR, but the distorted envelope is closer to the O~star at low $\beta$.  Also, the CWI wraps around the O~star, leading to polarimetric cancellation.  We use the \cite{canto_exact_1996} formalism to evaluate the possibilities.

\begin{figure}
\plotone{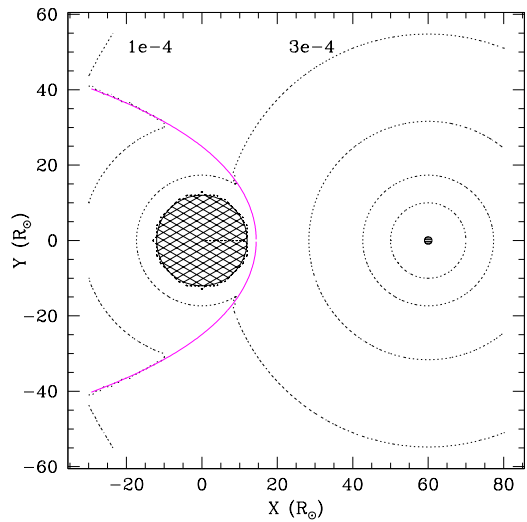}
\caption{Top-down view of a WR+OB binary with density contours superposed.  
The WR star is the smaller component at right; at left is the OB~star.  The magenta curve highlights the CWI shock.  Two representative contours indicate relative
densities, normalized to unity at the surface of the WR~star.  The contours are displayed in roughly 0.5 dex intervals.  In this example, the binary
separation is $60~R_\odot$, the OB star has a radius of $12~R_\odot$, and the WR star
has a radius of $1~R_\odot$.
\label{fig7}}
\end{figure}

To illustrate some of these features, Figure~\ref{fig7} shows a density contour plot in the orbital plane for a WR+OB binary.  The WR~star is the smaller star at right; the secondary is taken to be an O star.  The dotted curves are density contours at approximately 0.5 dex intervals, with two labeled for illustration, normalized to unit density at the surface of the WR~star.

To explore the polarization expected from WR+OB colliding winds, we adopted the following stellar and orbital parameters.
For the WR as primary and an O star as secondary, we assumed $M_1 = M_2 = 30~M_\odot$;
$R_1=R_\odot$ and $R_2 = 12~R_\odot$, $v_1=v_2$ so that $\alpha=1$; and we considered 
orbital scenarios ranging from a short-period orbit of $P_S=7$ d to a medium-period orbit of
$P_M=30$ d \citep[a typical range for colliding winds in circular orbits; see, e.g.,][]{fahed_colliding_2012, zhekov_x-rays_2012}.  Given the masses, these two orbits correspond to semi-major axes
of $a_S=60~R_\odot$ and $a_M = 160~R_\odot$. 

As a fiducial, we also adopted $\dot{M}_1= 10~\dot{M}_2$
so that $\beta=0.1$.  At this value of $\beta$, the relative standoff
distance for the bow shock is $R_{S,2}/D = 0.24$, which is a fixed ratio 
regardless of binary eccentricity, given that we assumed the winds are at
terminal speed.  For the short-period binary, we assumed a circular orbit, hence
$R_{S,2} = 14.4~R_\odot$. For a typical O~star wind, this would be well within the 
zone of wind acceleration, where effects such as radiative braking could be
significant \citep[as seen in the WR+O binary V444 Cyg;][]{lomax_v444_2015}.  For the sake of illustration, we ignored such effects.  

%
%
%
%
%

\begin{figure}
\plotone{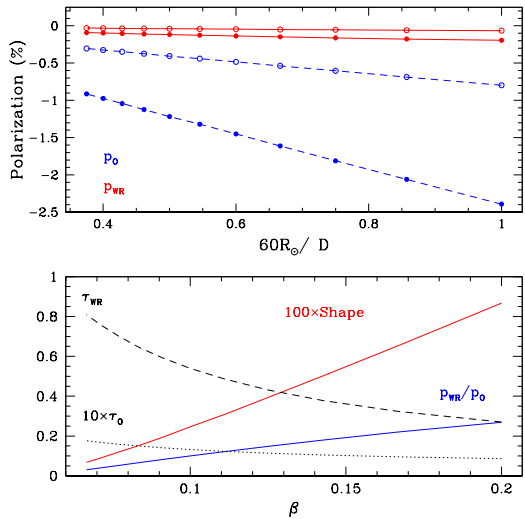}
\caption{Model results for a WR+O star binary (\S~\ref{sec:wr_ob_binaries}).  
The upper panel shows results for a fixed value of $\beta=0.1$ for the slow and fast wind
cases, plotted against binary separation as $D^{-1}$.  Blue lines represent $p_O$ and red lines
represent $p_{WR}$; solid circles are for slow models and open circles for fast wind
models.  The lower panel shows results for $D=160~R_\odot$ with $\beta$ varying
between 1/15 and 1/5.  Black curves represent the WR and O optical depths, as labeled.  
The blue line in this panel represents the ratio of $p_{WR}/p_O$.  The red line represents the ``shape,'' defined as the ratio
$(1-3\gamma_O)/(1-3\gamma_{WR})$ and scaled up by $100\times$.
\label{fig8}}
\end{figure}

We calculated $p_{WR}$ and $p_O$ for three scenarios, with a summary of results displayed in Figure~\ref{fig8}.  The first case is $\beta=0.1$ with orbital separation ranging from $60~R_\odot$ to $160~R_\odot$ for ``slow'' winds of 1000 km s$^{-1}$ for both stars.  The second case is for fast winds at 3000 km s$^{-1}$, with all other parameters fixed.  The third scenario corresponds to an intermediate wind speed of 2000 km s$^{-1}$ at a fixed separation of $D=160~R_\odot$, but with $\beta$ ranging between 1/15 and 1/5.

The upper panel of Figure~\ref{fig8} summarizes the comparison between slow and fast
winds.  Note that polarization is negative for our convention.  In this panel, the red lines represent $p_O$ and the blue lines represent $p_{WR}$.  The solid circles are for the slow wind cases, and the open circles are for the fast wind cases.  The results are plotted against $D^{-1}$, normalized as indicated.  The net result is that the polarization is overall larger for a slower wind, since the density is larger.  We find that $\tau_{WR}$ is roughly constant as $D$ changes, indicating that its value is dominated by the relatively extended spherical wind of the WR~star, since the CWI is far removed. Because the CWI is relatively farther from the WR~star with increasing $D$, $(1-3\gamma_{WR})$ becomes smaller with $D$.  Consequently, $p_{WR}$ decreases with increasing $D$.
The behavior for the O~star is that the polarization is dominated by the CWI.  The surface density of the CWI shock for the \citet{canto_exact_1996} solution scales as $D^{-1}$ overall.  This is
evidenced by the fact that both blue curves appear quite linear in the plot.

For the lower panel of Figure~\ref{fig8}, we display the results differently, as $\beta$ is allowed to vary between 1/15 and 1/5; with smaller $\beta$, the CWI is closer to the O~star component.  Consider first the dashed and dotted curves in black, for $\tau_{WR}$ and $\tau_O$, respectively.  As $\beta$ becomes smaller, $\tau_{WR}$ is larger, approaching the limit of the strictly spherical wind value.  The value of $\tau_O$ is much lower, and is plotted as scaled up by $10\times$.

The blue curve in this lower panel represents the ratio of $p_{WR}/p_O$. Its behavior indicates that from geometrical considerations, the contribution to the polarization from the O~star wind is much greater than for the WR~wind, even more so as $\beta$ becomes smaller.  Even though the WR~wind has a much higher optical depth scale, the distortion of the scattering envelope from spherical is quite minor  from the perspective of the WR~star.  This is made clear by the red curve, where ``shape'' is the ratio $(1-3\gamma_{WR})/(1-3\gamma_O)$, and scaled up by $100\times$.  From the perspective of the O~component, the scattering envelope is highly distorted.  

In combination, these results suggest that at wavelengths where the O~star is more luminous, the polarization will overall be larger (biased toward $p_O$) than at wavelengths where the WR~star is more luminous (polarization biased toward $p_{WR}$).  Our treatment does have limitations, the most important being that we ignore the wind acceleration region, and that we treat the WR~wind as optically thin to electron scattering. Radiative transfer models have shown that multiple scattering in bow shock structures can increase the degree of polarization as well as changing the polarization behavior with inclination angle \citep{Shrestha_2018}. However, in this case it is clear that the WR~wind is already a minor contributor to the polarization when $\beta \ll 1$, and a more full treatment of multiple scattering at the inner WR~wind is not expected to impact that conclusion. For the rare case of WR-WR binaries, multiple scattering could be significant and future modelling will need to take it into consideration. Inclusion of the wind acceleration region and associated density distribution, along with radiative inhibition, could certainly change the detailed outcomes of the models presented here.  Additionally, WR+O binaries can generally be expected to show chromatic behavior over a broader waveband than indicated in Section~\ref{sub:chromatic}.  Free-free opacity is important in the winds of WR~stars at all wavelengths; hence the WR stellar
spectrum is never Rayleigh-Taylor even though the OB SED can be \citep[e.g.,][]{1987ApJS...63..947H}.
Nonetheless, the present treatment indicates that $p_O \gg p_{WR}$, another qualitative result that is unlikely to change despite our more simplistic assumptions.

\subsection{Comment on the Overall Scale of the Polarization}

As mentioned above for identical stars, the polarization is proportional to ${\dot M}/v$.
This extends more generally to unequal stars, where the two different mass loss rates
receive different weightings, as a change in both mass-loss rates by any given factor
produces a change in polarization by that same factor.
This implies that a good model of the polarizing geometry allows the scale of the observed
polarization to yield a constraint on the stellar mass-loss rates that is
independent, and hence complementary, of all other methods for such determinations.
Furthermore, to the extent that the polarization is due to optically thin scattering,
the connection between polarization and mass-loss rate is independent of local clumping
in the winds.
Hence contrasting the mass-loss rates inferred from the overall scale of the polarization
to those obtained from conventional methods that are sensitive to clumping (often termed
``density-squared'' type mass-loss rate diagnostics such as radio free-free emission or
H $\alpha$ emission) provide an independent measure of the degree of clumping.
Constraining wind clumping is an important goal for understanding
the basic dynamics of radiatively driven winds.

\section{Summary}\label{sec:summary}

This study has made use of the theory of BME for optically thin electron scattering 
polarization for a massive colliding-wind binary.  The main novelty has been
to explore the analytic solution for the CWI shock structure in terms of shape and 
density from \citet{canto_exact_1996}, who assume radiative cooling to derive a thin shell result.  We assume axisymmetry throughout and explore polarization characteristics and contributions from
the two separate components.  Our results range from the limiting case of equal
star scenarios with $\beta=1$ and a planar shock interface (more appropriate to O+O binaries) to small $\beta$ scenarios (more appropriate to WR+O binaries).

Overall, there are numerous free parameters for the model, from the binary separation to the wind properties.  Even when most star and wind properties are held fixed, raising and lowering the terminal wind speeds at fixed ratio $\alpha$ still affects the polarization characteristics, since slower winds are more dense and faster ones are less dense.

Our three main results are as follows:

\begin{enumerate}

    \item From a detailed consideration of the contributions to the column density moments, there are various terms that scale with $D^{-1}$ for the binary separation.  However, for the equal wind scenario, the scale of polarization declines far less steeply than $D^{-1}$, so that even relatively wide binaries may display 
   a significant polarimetric amplitude, with a telltale orbitally varying phase angle.
   (See Fig.~\ref{fig3}.) 
    
    \item Chromatic effects can become quite significant toward UV wavelengths.  When dealing with massive stars, all of which are ``hot'' at $>10,000$~K, the optical emission is mostly or even very closely following the Rayleigh-Jeans law.  The consequence is that for optical and longer wavelengths, the continuum polarization is flat.  That polarization can still vary with orbital phase, but there are no chromatic effects.  However, at UV wavelengths for stars with different temperatures, the continuum polarization will generally deviate from flat (unless one star dominates the luminosity at all wavelengths).  The wavelength-dependent polarization provides additional diagnostic leverage for extracting information about the winds and CWI shock (Fig.~\ref{fig5}). This motivates UV polarimetric observations of colliding wind binaries, such as would be provided by the proposed {\em Polstar} satellite \citep{scowen_polstar_2021}.
    
    
    \item Orbital effects produce distinguishable shapes in the $q$--$u$ plane.  The shapes are mainly elliptical, as pointed out by BME already.  Importantly, we used the context of orbital effects to explore the influence of the CWI shock, and its boundary separating the two stellar winds, on the amplitude of polarization.  For this purpose we introduced the ``non-interacting winds'' (NIWs) construct.  This assumes an (unphysical) superposition of the respective two winds, with polarization arising solely from each star shining on the spherical wind of the other.  In this way the scenario for NIWs and CWIs can be compared on the scale of the same mass fluxes.  For the case of equal winds and a planar shock, inclusion of the CWI increases the polarization by factors of several, until the separation of the two stars becomes large compared to the stellar radii.  As expected, the CWI and NIW polarizations become equal, since the CWI is far removed from either star and thus adds only a small column density compared to the spherical winds.
    
    
    
    \item When $\beta \ll 1$, as for example in the case of WR+OB binaries, we find the interesting result that the polarization for the OB~component is much higher than for the WR~component.  Whether the observed polarization is dominated by the WR or the OB star will depend on the weighting by the specific luminosities.  However, it is clear that at wavelengths where the WR star is more luminous, the polarization will be lower as set by $p_{WR}$, and where it is less luminous, the polarization will be higher as set by $p_O$ (or $p_B$, as the case may be; Fig.~\ref{fig8}.)
    
\end{enumerate}

In closing it is worth noting that the individual stars in a massive
colliding wind system may themselves be sources of polarization, which
may be steady or variable.  For example, around 10\% of massive stars
are known to be magnetic \citep{2016MNRAS.456....2W}, and it is possible
(although very rare)
for massive star binaries to have a component that possesses a significant
magnetic field \citep[e.g., Plaskett's Star,][]{2013MNRAS.428.1686G}.
\cite{2022MNRAS.511.3228M} has recently explored the effects of variable
linear polarization from electron scattering for rotating magnetospheres.
While this could complicate efforts to isolate the variable polarization
from the CWI, the polarization from individual stars will be modulated
on a rotation period whereas the colliding wind polarization is modulated
on the orbital period.  Unless the binaries are very close, these periods are
unlikely to be the same.

WR stars in particular are known to be sources of polarization.  However,
typical polarization behavior from individual WR stars appears
stochastic \citep[e.g.,][]{1987ApJ...322..870S, 1987ApJ...322..888D}.
The behavior is likely associated with the wind flow time,
$R_\ast/v_\infty$, that is much shorter than binary orbital periods.
In addition to being stochastic in nature, the effect could be averaged
out to emphasize the smoother variable polarization from the CWI
on the longer period of the orbit.  In addition to variable
polarization, some WR~stars may have long-term stable polarizations
\citep[e.g.,][]{1998MNRAS.296.1072H}.  \cite{2022A&A...658A..46A}
have explored the polarization that could result for axisymmetric
rotationally distorted winds of WR~stars.  However, such polarization
would be constant.  The effect would be to contribute to a constant
offset to the system polarization, similar to the effect of
interstellar polarization.  Variable polarization would arise entirely
from the CWI over the timescale of the orbital period. 

\acknowledgments

RI acknowledges funding support
for this research from grants by the National Science Foundation (NSF),
AST-2009412 and AST-1747658.
 YN acknowledges support from the Fonds National de la Recherche Scientifique (Belgium), the European Space Agency (ESA) and the Belgian Federal Science Policy Office (BELSPO) in the framework of the PRODEX Programme. JLH is grateful for NSF funding under award AST-1816944, and acknowledges that the University of Denver occupies land within the traditional territories of the Arapaho, Cheyenne, and Ute peoples.  NSL wishes to thank the National Sciences and Engineering Council (NSERC) for financial support.

\appendix

\section*{Special Case of $\alpha=\beta=1$}\label{AppA}

When $\beta=1$ and $\alpha=1$, with stars of identical stellar and
wind parameters, the CWI is planar, and the
solution for the shock properties simplifies considerably.
First, we introduce $\theta=\theta_1=\theta_2$ as the angle from either
star to a point on the planar shock.  The distance of the shock from
either star becomes 

\begin{equation}
R_S = D/2\mu.
\end{equation}

\noindent The projection factors become $K=K_1=K_2 = \mu$.
Simplification of the surface density was noted already in eq.~(\ref{eq:planarSigma}).
Contributions to the polarization from the CWI component depends on the following integrals (see eqs.~[\ref{eq:A}]-[\ref{eq:D}]):

\begin{equation}
\int_0^1 \Sigma(\mu)\,K(\mu)\,d\mu =
	4\Sigma_0\,\int_0^1 \frac{\mu^2(1-\mu)^2}{\sin\theta(\theta
	-\mu\sin\theta)}\,d\mu \approx 0.56\,\Sigma_0,
\end{equation}

\noindent and

\begin{equation}
\int_0^1 \Sigma(\mu)\,K(\mu)\,\mu^2\,d\mu =
	4\Sigma_0\,\int_0^1 \frac{\mu^4(1-\mu)^2}{\sin\theta(\theta
        -\mu\sin\theta)}\,d\mu \approx 0.32\,\Sigma_0.
\end{equation}

Using these two results, we can analytically derive the polarization for
the scenario of two equal stars that are widely separated, with $D \gg R_\ast$.
At wide separation we can ignore the finite size of each star (i.e., $\theta_\ast\rightarrow 0$), which amounts to not having to consider case C (c.f., \S~\ref{sub:caseC}).
Additionally, the region of case A for each star (c.f., \S~\ref{sub:caseA}) is hemispherical and consequently makes no contribution to the net polarization.  All that remains are contributions I., II., and III. for case B (c.f., \S~\ref{sub:caseB}), where the limits of the angular integrations are 0 to $\pi/2$ in $\theta$ or $+1$ to 0 in $\mu$.  The total polarization becomes:

\begin{eqnarray}
p & = & \frac{3}{8}\,\sigma_T\,\left( \bar{N} - 3\tilde{N}   \right) \\
  & = & \frac{3}{8}\,\sigma_T\,\left[ \bar{N}_{B,I}+\bar{N}_{B,II}+\bar{N}_{B,III} - 3\tilde{N}_{B,I} - 3\tilde{N}_{B,II} - 3\tilde{N}_{B,III} \right] \label{eq:terms} \\
  & = & \frac{3}{8}\,\sigma_T\,n_0\,R_\ast\,\left(\frac{R_\ast}{D}\right)\,\left[ -1.00 + 0.56 +2.47 +075 -0.96 -6.30    \right] \label{eq:numbers}\\
   & = & -1.68\,\sigma_T\,n_0\,R_\ast\,\left(\frac{R_\ast}{D}\right), \label{eq:pCWI}
\end{eqnarray}

\noindent where each number in the square brackets of eq.~(\ref{eq:numbers}) corresponds to
each term in the preceding line of eq.~(\ref{eq:terms}).  Note that if grouped by region,
each of I., II., and III. would separately yield net negative polarizations.

For comparison the polarization for the non-interacting wind (NIW) case introduced in \S~\ref{sub:orbital} can also be evaluated analytically for
$D \gg R_\ast$.  The two relevant angle-averaged column densities are

\begin{equation}
\bar{N} = n_0\,R_\ast\,\left(\frac{R_\ast}{D}\right)\,\int_0^{\pi/2} \pi\,d\theta   = \frac{\pi^2}{2}\,n_0\,R_\ast\,\left(\frac{R_\ast}{D}\right),
\end{equation}

\noindent and 

\begin{equation}
\tilde{N} = n_0\,R_\ast\,\left(\frac{R_\ast}{D}\right)\,\int_0^{\pi/2} \pi\,\cos^2 \theta\,d\theta   = \frac{\pi^2}{4}\,n_0\,R_\ast\,\left(\frac{R_\ast}{D}\right).
\end{equation}

\noindent For an NIW with wide binary separation, the polarization is 

\begin{equation}
p_{NIW} = \frac{3}{8}\,\sigma_T\,n_0\,R_\ast\,\left(\frac{R_\ast}{D}\right)\,\left(\frac{\pi^2}{2} - 3\frac{\pi^2}{4}\right) = -0.93\,\sigma_T\,n_0\,R_\ast\,\left(\frac{R_\ast}{D}\right).
\label{eq:pNIW}
\end{equation}

\noindent The ratio of the coefficients from eqs.~(\ref{eq:pCWI}) and (\ref{eq:pNIW})
is 1.8, which is the solid green line appearing in Fig.~\ref{figorb}.

\bibliography{CWB_bib}

\end{document}